# Comparative analysis of recirculating and collimating cesium ovens

Raphaël Hahn  ; Thomas Battard; Oscar Boucher; Yan J. Picard; Hans Lignier; Daniel Comparat   ;
Nolwenn-Amandine Keriel; Colin Lopez; Emanuel Oswald; Morgan Reveillard  ; Matthieu Viteau

View Online

Export Citation

CrossMark

---

**Articles You May Be Interested In**

Candlestick oven with a silica wick provides an intense collimated cesium atomic beam

*Rev Sci Instrum* (February 2007)

Recirculating atomic beam oven

*Rev Sci Instrum* (August 2008)

Effusive atomic oven nozzle design using an aligned microcapillary array

*Rev Sci Instrum* (February 2015)

31 August 2023 09:18:24

AIP Publishing

# Comparative analysis of recirculating and collimating cesium ovens

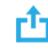

View Online    Export Citation    CrossMark

Raphaël Hahn,[1,a)]  Thomas Battard,[1] Oscar Boucher,[1] Yan J. Picard,[1] Hans Lignier,[1] Daniel Comparat,[1,b)]
Nolwenn-Amandine Keriel,[2] Colin Lopez,[2] Emanuel Oswald,[3] Morgan Reveillard,[4]  and Matthieu Viteau[4]

### AFFILIATIONS

[1] Université Paris-Saclay, CNRS, Laboratoire Aimé Cotton, 91405 Orsay, France
[2] Université Paris-Saclay, ENS Paris-Saclay, CNRS, LuMIn, 91190 Gif-sur-Yvette, France
[3] Physics Department, CERN, 1211 Geneva 23, Switzerland
[4] Orsay Physics, ZAC ST Charles, 3ème Avenue, No. 95, 13710 Fuveau, France

a) **Current address:** Laboratory of Physical Chemistry, ETH Zurich, CH-8093 Zurich, Switzerland.
b) **Author to whom correspondence should be addressed:** daniel.comparat@universite-paris-saclay.fr

### ABSTRACT

We have performed a study of several cesium oven designs. A comparison between recirculating (or sticking-wall) and collimating (or re-emitting-wall) ovens is made in order to extract the most efficient design in terms of beam brightness. Unfortunately, non-reproducible behaviors have been observed, and the most often observed output flux is similar to the sticking-wall case, which is the lowest theoretical value of the two cases, with a beam brightness close to $10^{18}$ at. $sr^{-1}$ $s^{-1}$ $cm^{-2}$. The reason of this universally observed behavior is unclear despite having tested several materials for the collimating tube. Conclusion on possible improved design based on sticking of cesium on several (un)cleaned surfaces is given.

Published under an exclusive license by AIP Publishing. https://doi.org/10.1063/5.0085838

## I. INTRODUCTION

Collimated atomic beams have applications in atomic physics, chemical physics, and surface sciences. Unfortunately, even if the generation of intense beams using atomic ovens has a long history, practical information on various aspects of oven design[1–4] and a comparison between theory and experiment or a comparison between different experimental datasets are still incomplete.[2,5–10] For instance, as quoted in Ref. 2, no comparative study exists concerning "the advantages and disadvantages of recirculating ovens with ovens that use a collimating array to produce a beam."

In our group, we have been interested in designing an intense collimated atomic cesium beam for various atomic physics studies. One goal being to produce an intense ion or electron beam once ionized.[11–14] Similar goals for other atoms are pursued in many groups, for instance, Ca,[15] Sr,[16,17] Yb,[18] Ba,[19] Li,[4] and Rb.[20] A systematic study on any atom might thus be useful for the atomic physics community as a whole.

Cesium beams, in particular, have many applications such as fundamental frequency standard studies, Global Positioning System (GPS) constellation,[21,22] or the production of negative ions (Cs reduces the work function of the surfaces on which it is deposited)[23–28] extensively used in plasma fusion installations.[29–33]

As detailed later, finding the best Cs oven design from the literature was not so obvious. The only clear statement is that a high partial pressure in the oven produces a high flow rate.[2] However, the emittance of the beam also increases when rising pressure, and the regime evolves from molecular to supersonic. The performances of an atomic beam are expected to be determined by geometrical characteristics (tube length, aperture diameter, etc.) or physical variables (temperature, pressure, etc.).[34] However, discrepancies between the calculated flux and measured flux of cesium ovens, common in the literature,[35] indicate that other factors can be preponderant. Indeed, the usual flux calculations consider no chemical interaction between emitted atoms and the surface of the tube. In the case of cesium, because of its high reactivity, this hypothesis is unlikely to be correct.

We thus had strong incentive to studying interactions between cesium and a tube to collimate or (re-)emit cesium atoms. We shall see that physical-chemistry parameters play an important role in the

31 August 2023 09:18:24

performance of the oven: the nature of the tube material, its cleanliness, or its surface state. In this article, we shall thus present our conclusions based on our decennial work to design a simple atomic cesium beam with low emittance, high flux, and low consumption (long lifetime) and being compatible with ultrahigh vacuum (meaning low pumping speed requirement). We believe that this kind of testimony can be interesting for the community. Our goal is not to provide an extensive and detailed study of all beam parameters such as the velocity or spatial distribution at different distances from the nozzle output but to provide results such as an on-axis flux useful enough for practical application.

This article is therefore organized as follows: we first recall the basic theoretical tools and the geometrical considerations that help in the design of ovens with or without collimation tubes. We then review the literature on cesium ovens and compare the experimental properties with theoretically expected ones. We then present some of our experimental designs and results and also expose the evolution of our experimental setup to a more and more efficient system by studying several designs of recirculating or collimated ovens with a tube and changing the orientation (vertically/horizontally) of the atomic beam. We compare the output flux and how easy these ovens can be handled. Finally, a discussion on the choice of the material and of the cleanliness of the surface is given.

## II. EXPECTATION FOR EFFUSIVE OVENS

It is beyond the scope of this paper to review the theory of atomic ovens that can be found in standard textbooks.[2,7–10] We simply recall here few important aspects. After the work of Clausing, it was quickly recognized that a narrower and more pointed beam pattern emerged from a tube as the ratio $L/d$, between the length of the tube $L$ and its diameter $d$, increases. Thus, we shall only study cases of the so-called "long tubes" where $d \ll L$, and most examples presented here will have an aspect ratio of $L/d \sim 50$.

Different regimes occur depending on the value of the mean free path of the gas $\lambda = \frac{k_B T}{\sqrt{2}\pi P \sigma^2}$, where $P$ is the pressure, $T$ is the temperature, $k_B$ is Boltzmann's constant, and $\sigma = 0.27$ nm is the kinetic diameter.[36–38] In a Cs vapor, we have $\lambda = 10$ cm, 1 cm, 1 mm, 0.1 mm, and 10 $\mu$m for T $\approx$ 90, 130, 180, 240, and 330 °C, respectively.

In an oven at temperature $T$, the pressure is the saturated vapor pressure $P$, and so, the atomic density is $n = P/k_B T$. For cesium, the vapor pressure is given by the empirical formula $P = 1.4825 \times 10^9$ Pa $\times e^{-\frac{0.78\,eV}{k_B T}}$.[39] Because of the two different physical length scales (d and $L$), there are two different Knudsen numbers (Kn = $\lambda/d$ and Kn = $\lambda/L$) to determine the regime. The so-called molecular, transparent, or diffusive regime occurs when $d < L < \lambda$, the intermediate or opaque regime occurs when $d < \lambda < L$, and the collisional, viscous, hydrodynamical, or continuum regime occurs when $\lambda < d < L$, which can evolve to the supersonic regime at high pressure (Knudsen numbers smaller than unity).

One problem with oven design is that, in practice, they are often operated in an intermediate regime while they are conceived for the transparent regime[3] where equations are simpler and often (not in our cesium case as we shall see) agree with experiment.[2] The transparent regime mode of operation physically means that atoms can enter and exit the tube without being subjected to any inter-atomic collisions but can still interact with the oven walls. These atom–wall collisions can lead to bonding of the atom to the wall (sticky collision) or to quick re-emission of the atom (non-sticky collision).[33] At the studied oven-temperatures, the proportion of cesium dimers in the atomic beam is less than 0.1%.[40] Furthermore, heating the tube (nozzle) reduces the dimer fraction.[9,41,42] We will thus neglect the dimer fraction in this first description of oven properties. However, dimer formation should not be forgotten as they can be efficient seeds for further nucleation as we will see later.

Many types of ovens exist, and we will not review all of them. Because the terminology does not seem to be established very well, we would like to clarify some basic terms (see Fig. 1). Conventional ovens use long narrow tube(s) to achieve good collimation. If the tube re-emits atoms that hit its wall, the oven is called a "bright-wall" oven (or tube-collimated ovens)[2,43] [cf. Fig. 1(a)]. If the tube "catches" the atoms that hit its wall, then only atoms that are emitted in a straight line from the oven go through the tube: it is then a "dark-wall" oven. In this last case, the beam output (in transparent regime) is identical to the one produced by two apertures (or slits) at the entrance and the output of the collimation region [cf. Fig. 1(b)]. This "two-aperture" concept can be used with simple designs [such as in Fig. 1(b)]. We will call a "recirculating oven" such an oven without any collimating tube but with only an aperture and a mechanism that returns the non-emitted atoms to the source. A recirculating oven can also be achieved, for instance, by capillary action using a porous tube[43] or simply using gravity [see Fig. 1(d)]. In this work, we will model only two types of ovens: the "bright-wall, tube-collimated" model that re-emits the atoms from a collimating tube and the "dark-wall" model that considers only atoms that go straight through and that can be made of a sticking tube or of two apertures with possible re-circulation. During the course of this study, we have used all the types of ovens presented in Fig. 1.

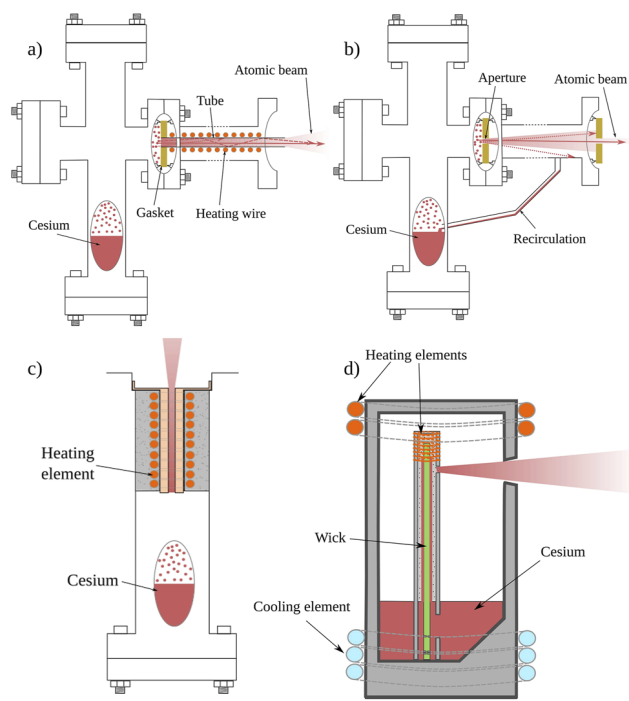

**FIG. 1.** All types of Cs ovens studied in this work: (a) horizontally tube-collimated, (b) recirculating,[44] (c) vertically collimated, and (d) candlestick ovens.[45] Dark-wall (bright-wall) behavior is represented in (a) and (b) by dotted (dashed) lines.

In the non-viscous regime, the theory of the output beam is well established.[2,46] With $\dot{N}$ the total number of atoms flowing through the orifice per second, we have $I(\theta)d\omega$ the number of atoms flowing through it per second into a solid angle $d\omega = 2\pi \sin(\theta)d\theta$ in a direction making an angle between $\theta$ and $\theta + d\theta$ with the normal to the area and we have $\dot{N} = \int I(\theta)d\omega$.

Atoms that go through the tube can be emitted directly from the first orifice or from the surface of the tube. From the first orifice, the flux $I(\theta)$ that goes out of it noted $I_0(\theta)$ is given by[9]

$$I_0(\theta) = \frac{dn\bar{v}}{4\pi}\frac{d}{2}\cos\theta\left[\arccos q - q\sqrt{1-q^2}\right],$$

with $q = L\tan\theta/d$, $\bar{v} = \sqrt{8k_BT/\pi m}$ being the atom's average velocity, and $m$ being the mass of a single atom. The standard theory uses a cosine (Lambert) re-emission angular distribution, often justified by the simple assumption that atoms colliding with the wall lose track of their original motion because of the arbitrary microscopic-scale aspect of the surface or because of adsorption prior to re-emission.

Even if this Lambertian cosine angle emission is found to be not fully correct, we will use it here for its simplicity.[47–49] Under this assumption,[9,43,46] the flux $I(\theta)$ coming from the tubes arises from atoms re-emitted by the tube-walls is noted $I_{\text{tube}}(\theta)$ and is

$$I_{\text{tube}}(\theta) = \int_0^{\min(L,d/\tan\theta)} \frac{dn\bar{v}}{4\pi} \sin\theta\sqrt{1 - \left(\frac{z\tan\theta}{d}\right)^2}\,dz.$$

$n(z)$ and $\bar{v}(z)$ can vary along the tube (for instance, due to the local temperature). Therefore, the atomic flux is given by

$$\dot{N} = \frac{n\bar{v}\pi d^2}{16}W = I(0)W, \qquad (1)$$

where $W = \int I(\theta)/I_0(\theta)d\omega$ is called the transmission probability (or the Clausing factor). Previous formulas and basic geometrical consideration lead to a transmission probability

$$W_{DW} = (d/L)^2 \qquad (2)$$

for a pure dark-wall (DW) where $I(\theta) = I_0(\theta)$ and a recirculating oven (or similarly with a sticking tube) and

$$W_{BW} = 4d/3L \qquad (3)$$

for a non-sticking bright-wall (BW) tube where $I(\theta) = I_0(\theta) + I_{\text{tube}}(\theta)$. Thus, for typical values of $\frac{L}{d} \approx 50$, we have $\frac{W_{BW}}{W_{DW}} \approx 67$.

As mentioned previously, another important aspect is the beam divergence (FWHM angular width) $2\theta_0$, defined by $I(\theta_0) = I(0)/2$ and the on-axis intensity $I(0)$. In the transparent regime, the angular

distribution of the atomic beam differs only slightly between the two cases with

$$2\theta_0 = 1.7d/L \qquad (4)$$

for a bright-wall collimated beam[2] and

$$2\theta_0 = d/L \qquad (5)$$

for a dark-wall recirculating oven (from simple ray-tracing consideration).[2] The beam profile deteriorates similarly in both cases with increasing backing pressure because of the increase in gas–wall and gas–gas collisions,[51] leading to a square-root dependence of $\theta_0$ and $I(0)$ on $P$ at high pressure.[2] For instance, with $T$ in Kelvin, $L$ in mm, $P$ in Pascal, and $\sigma$ in pm, the full divergence angle is given by[2]

$$2\theta_0 = 1.7\frac{d}{L}\frac{1}{\text{erf}[76.4\sqrt{T}/(\sigma\sqrt{LP})]}. \qquad (6)$$

The parameters (flux, divergence, density, and peaking factor) to be optimized are always experiment-dependent, and there are several possible parameters to describe the shape of an atomic beam such as $\theta_0$ linked to the atomic density $\rho = I(0)/R^2$ at a distance $R$ or the peaking factor $\kappa = \pi I(0)/\dot{N} = \pi/W$ that represents the improvement of the beam over the one produced by an ideal single orifice. An interesting parameter to characterize the beam quality is the beam brightness $B$, which is the total flux $f = \dot{N}$ divided by the emitting area $A = \pi d^2/4$ and by the solid angle divergence that, for the small angles we have here, is $\Omega = \pi\theta_0^2$:[52–54]

$$B = \frac{f}{A\Omega} = \frac{\dot{N}}{(\pi d^2/4)(\pi\theta_0^2)} = \frac{n\bar{v}W}{4\pi\theta_0^2}. \qquad (7)$$

We thus expect $B_{DW} = \frac{n\bar{v}}{\pi}$ for a dark-wall recirculating oven and $B_{BW} = 0.46\frac{L}{d}B_{DW}$ for a bright-wall oven. For typical values of the $\frac{L}{d} \approx 50$, we have $\frac{B_{BW}}{B_{DW}} \approx 23$.

To summarize, the main message from theory is that to get an intense collimated beam, a bright-tube-collimated oven seems far superior than a recirculating one, and a high aspect ratio $L/d$ is favorable.

## III. EXPERIMENTAL DESIGNS

A small ($d$ or $L$) geometry is favorable to avoid entering the hydrodynamical or even the opaque regime. Thus, multi-array tubes should be much better than a single "big tube" with a similar opening area because of possible much higher pressure still being in the molecular regime.[55–57] Therefore, micrometer diameter capillary arrays (or even nanometer diameter aligned carbon nanotubes) seem ideal. Thus, in order to minimize the effects of inter-atomic scattering while maintaining a minimal angular spread and a high beam intensity, several groups have implemented nozzles consisting of numerous microchannels with diameters ranging from a few micrometers to a few hundred micrometers[4] sometimes using conical geometry to enhance the density.[58] Our ovens and their characteristics in terms of temperature, types (mainly with tubes, capillaries, or slits), flux, flux per unit area and per solid angle, and brightness is given as the Appendix to this article in Table I. Detailed experimental studies of several types of microtube-ovens (parallel capillary arrays, focusing capillary arrays, and variable density rings) have been performed with different gases, and a reasonably good agreement was obtained between theory and experiments.[19,34,59] For cesium, however, the situation does not seem to be so clear as it is shown in Fig. 2 and in Table I where we have plotted the experimental flux and the brightness for most of published Cs ovens. The results are extremely diverse with no simple link between flux, brightness, and oven design. One of the main reasons for this diversity is that achieving an actual bright-wall behavior with cesium is very difficult. On the contrary, it is easy to produce a dark-wall by cooling the tube so that the Cs atoms stick on it[59,60] or simply by using a two-aperture configuration [see Fig. 1(b) and Table I]. Achieving a bright-wall tube requires that the nozzle (tube) temperature should be at least 50 °C higher than the temperature of the crucible to avoid clogging,[1] but extensive experimental data demonstrate that this is not a sufficient condition. Furthermore, several types of nozzles (tubes) have been used, and thus, a comparison between them is not straightforward.

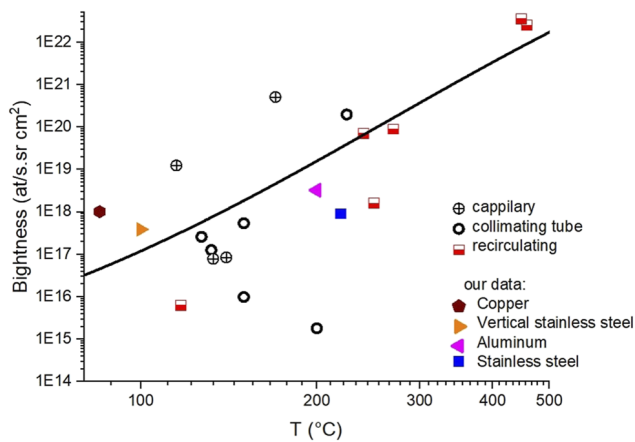

**FIG. 2.** Experimental flux per unit of area and solid angle (brightness in at. s⁻¹ cm⁻² s⁻¹) for several Cs ovens found in the literature compared to the theoretical expression for a dark-wall oven of Eq. (7) (solid line). The two highest temperature points are in the supersonic regime.

The observed dispersion of results (see Fig. 2 and Table I) was the main motivation for our study. As discussed previously, a popular approach to design atomic and molecular beams is to use stacks of multiple tubes. However, in order to simplify the study, we focus here on a single tube and try to find the best design with a single tube. It will then be possible to multiply this tube choice in a multi-channel design in order to be able to increase the pressure (and still being in a molecular regime where the mean free path is the bigger dimension).

## IV. OUR STUDY

We have performed many tests based on a single tube or simple recirculating ovens with the different designs shown in Fig. 1. Most of our ovens were operated between 300 and 450 K. In the high temperature range, the opaque regime starts to play a dominant role, the flux saturates, and the beam divergence increases.

## A. Detection methods

Several detection methods of the Cs flux exist (quartz balance, Auger peak height,[64] and Langmuir–Taylor surface ionization hot wire detector). Here, we used a light absorption method based on the Beer–Lambert–Bouguer law. In most of our studies, we record the absorption spectrum when scanning the frequency of a low intensity diode laser (some details can be found in several references such as Ref. 76). This gives both the spatial and velocity profile. Indeed, the beam divergence can be studied using the Doppler effect, and the beam profile can be measured simply by a spatial scan of the probe beam. We used a power of a few micro-watts collimated on a few hundred micrometers to measure the absorption of the $6s(F = 4) \rightarrow 6p_{3/2}(F' = 5)$ closed hyperfine transition that is proportional to the line integration of the atomic density.

## B. Recirculating oven

As stated in Ref. 4, recirculating oven designs are complex and offer little or no increase in lifetime over channel designed ovens. Indeed, our first recirculating oven was the candlestick design [Fig. 1(d)[45]], and we found it very tricky to handle.[83] Therefore, we quickly moved to a simple but very robust recirculating design [see Fig. 1(b)] inspired by Ref. 44. For these ovens, the only care is to avoid accumulation of cesium in some parts that can then create a large vapor in addition to the desired beam. The flux and the angular distribution of such recirculating ovens follow very well the recirculating dark-wall theory in the transparent regime. An example of results is shown in Fig. 3(b).

## C. Collimating tube

When going from a dark-wall (recirculating) oven [see Fig. 1(b)] to a collimating-heated-tube to produce a bright-wall oven [see Fig. 1(a)], we expected to observe higher flux, as predicted by Eqs. (2) and (3). The gain $\frac{W_{BW}}{W_{DW}}$ is given by the ratio of the transmission probabilities and is thus of the order of the aspect ratio $L/d$. However, the choice of the best possible material for the tube (but also for the oven crucible) is far from being obvious and might have strong influence on the dark-/bright-wall behavior. Indeed, looking at the list of Cs ovens in Fig. 2 (and in Table I), all possible results arise and some published ovens even have flux higher than predicted by the theory (Ref. 64, for instance). It is thus interesting to go back to theory to try to get some insight into the material choice and see how this might influence the oven properties.

### 1. Theory of interaction of Cs atoms on surfaces

Atom–wall collision processes, a subset of surface interaction science, are ubiquitous in nature and yet are far from being well understood.[84] Chemical dynamics at the gas–surface interface can be very complex[85,86] with elastic or inelastic scattering, trapping-desorption, chemical reactions at surfaces, and thus possible adsorption of vapor or gas on the tube-walls or surface diffusion.[87] Basic processes of atom–surface interactions are physical adsorption (physisorption) due to van der Walls interactions and chemical adsorption (chemisorption) due to the formation of a chemical bond with the wall atoms. This can be followed by desorption, which is much more probable in the physisorption case. For alkali metals, the physics and chemistry are already complex,[88,89] and it is even more pronounced with a highly reactive atom such as Cs. Furthermore, the interaction of cesium with a totally or partially cesiated surface adds extra complexity for instance with surface bounding on defects, and fractal formation.[90] For instance, the mechanism of cesium transport through tungsten tubes has been studied and indicates that cesium atoms can condense because of being held temporarily, but they can also jump from its landing place to other sites,[91] leading to surface diffusion, molecular flow, and ionic flow under the influence of electric fields, which exist in the tube.[92]

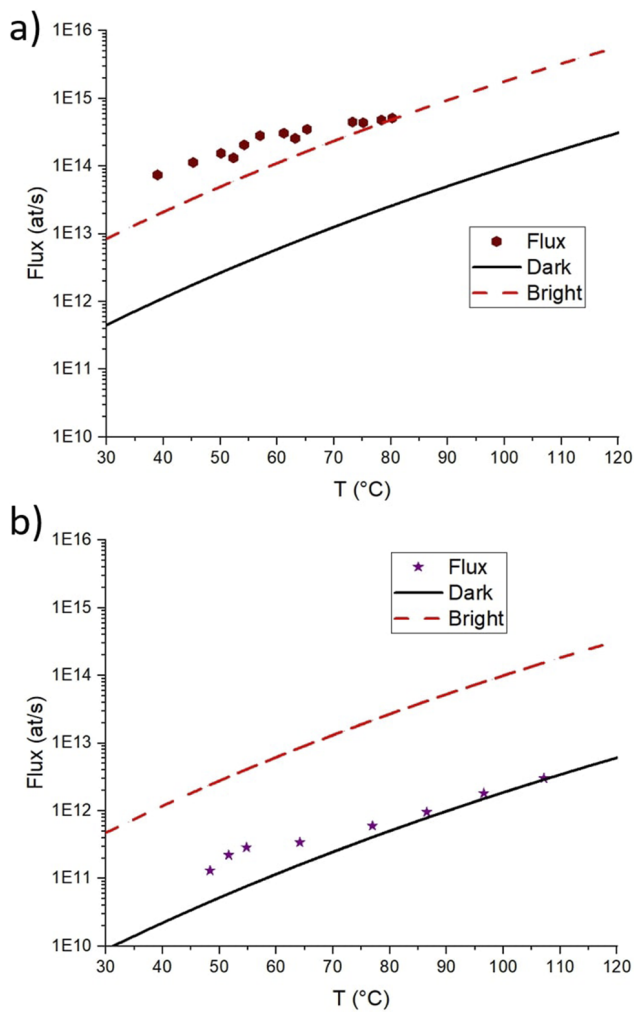

**FIG. 3.** Experimental vs theory for two type of ovens. (a) Bright-wall oven [Fig. 1(a)] with $d = 5$ mm and $L = 70$ mm. (b) Dark-wall oven [Fig. 1(b)] with $d = 2$ mm and $L = 80$ mm. In both cases, the theory is given for the dark-wall from Eq. (2) (black line) and the bright-wall from Eq. (3) (dashed-red line).

All this illustrates the very high complexity of the system. Nevertheless, we can try to be guided by simple models to try to rationalize what should be a bright-wall behavior. A simple expression for the adsorption time (or desorption rate) of an atom on a surface at temperature $T$ is $\tau = \tau_0 e^{E_a/k_B T}$ with $E_a > 0$ the so-called adsorption energy (binding energy of the atom–surface potential curve) and $\tau_0$ is a characteristic time (such as the period of vibration of the adsorbed atom in the wall potential): $\tau_0 \sim h/k_B T$.[85] From the $\tau$ value, the surface coverage is given by $\theta = n_S/n_0$ where $n_0 = 5.2 \times 10^{14}$ at.cm$^{-2}$ is the Cs surface density for a complete monolayer[31] and $n_S = \frac{n\bar{v}\tau}{4}$ with $n$ being the local vapor Cs gas density. Typical values are $E_a \approx 0.5$ eV and $\tau_0$ in the picosecond range, leading to $\tau$ near the low microsecond range (at 400–500 K) and a surface coverage in the percent range. This model thus does not predict the formation of multilayers. The simple idea of choosing a material with small $E_a$, where individual Cs atoms have a low probability of forming a lasting bond, seems to be the way to go to produce a bright-wall oven. Unfortunately, the parameters of the models are often hard to derive and may deviate from article to article quite a lot and even by many orders of magnitude for $\tau_0$.[77,85,93] Furthermore, this one-particle type of model is very simplified and does not accurately describe the behavior of Cs sticking. For example, if a Cs atom approaches the surface where already adsorbed Cs atoms are present, they might form a more stable compound, which will lead to the accumulation of Cs. Thus, adsorbate–adsorbate interactions have to be included. These interactions strongly depend on the temperature, and above a fraction of monolayer (a coverage $\theta < 0.1$ typically), surface-clustering and coalescence occur.[94–97]

A better theory to estimate the number of multilayers is the BET (Brunauer–Emmett–Teller) model.[98,99] It links the surface coverage $\theta$ to the ratio $p = P/P_W$ between the vapor pressure $P$ in the oven crucible (so at the temperature of the oven's reservoir $T$) and the saturating vapor pressure $P_W$ at the wall surface temperature $T_W$: $\theta = cp\frac{1}{1-p}\frac{1}{1+(c-1)p}$ with $c = e^{(E_a - E_c)/k_B T_W}$, where $E_c$ is the heat of vaporization per atom (a positive value, equal to minus the heat of condensation), which is, in this model, assumed to be the heat of adsorption for the second and higher layers. This indicates that heating the walls by increasing $P_W$ is indeed required to reduce $p = \frac{P}{P_W}$ and so the formation of multilayers ($\theta \approx cp$ becomes small if $p$ is small). However, the large heat of vaporation of Cs ($E_c \approx 0.8$ eV[100]) tends to lead to the formation of multilayers.

The main conclusion from the theory is that we should avoid material with high adsorption energy with Cs, such as metals, because of the charge-image that will attract the atom to the surface through the van der Waals force. The adsorption energy of Cs on Cu, for example, is $\approx 2.45$ eV[101] even if Cs is very mobile. The expectation of a high adsorption energy (eV range) of cesium on metals is also supported by theory[102] including detailed theoretical works based on density functional theory calculations (DFT) that help to elucidate the properties of alkali metals when supported on metal and oxide surfaces.[101] However, contradictory results have also been reported, such as the fact that the condensation coefficient of cesium on nickel was found to be 0.1.[103] A final problem with metal is that liquid cesium dissolves most of the metallic elements even if it seems possible to use, for instance, Ti, Zr, V, Ta, Mo, or W.[49] Therefore, on metals or on other surfaces, it is not obvious to conclude about the (multi)layer formation. It is thus not easy to choose a material that realizes for sure the initial idea of Cs "rebound" on the tube surface to realize a bright-wall oven.

These materials should thus be tested as tube-materials but keeping in mind that these surface studies are typically done under very high cleanliness and ultrahigh vacuum conditions, whereas our ovens operate at non-negligible cesium vapor pressure. This means that the behavior of Cs on these materials might very well differ from what is observed in these studies. However, other kinds of experiments such as those on vapor cells are typically conducted at higher pressure, leading to other kinds of potentially interesting materials.

### 2. Material based on Cs coated cell experiments

Stored Cs beams in a cell have been studied as early as in the 1960s by Ramsey and collaborators.[104] This first study shows that paraffin works (as expected because $E_a \sim 0.1$ eV[105]) but wax, Teflon, LiF, sapphire, glass, fused quartz, and nickel adsorbed the beam and polyethylene and pyrolytic graphite gave not reproducible (but sometimes promising) results due to the difficulty in preparing a clean surface. After this, numerous studies of coatings to prevent spin-relaxation of alkali-metal atoms have been performed, such as the detailed study of Ref. 93. A first promising candidate for our study can be alumina ($Al_2O_3$[89] or properly clean sapphire[93,106]) because alumina coating was found to improve the resistance of a vapor cell by a factor 100[107] and also because, if above $300\,^{\circ}C$ Cs attacks glass, this effect is limited for $Al_2O_3$.[45] Unfortunately, strong interactions were also observed such as ionic chemisorption of Cs on $Al_2O_3$ at surface defect sites (steps, kinks, and corners)[108] or Cs adsorption due to interaction with the unsaturated oxygen surface such as s–p hybridation between the s valence orbital of the alkali atoms and the 2p oxygen orbital.[109,110]

A second type of promising surface is the H-terminated surface that is believed to not adsorb any Cs atom.[111] Therefore, hydrogen termination of CVD diamond films (or DLC: diamond-like carbon) might be very promising.[96,112] Similarly, long methyl chains protect the surfaces from Cs attachment.[93,106]

However, here, again adsorption and desorption models do not agree well with coating experiments.[90,113] Therefore, as summarized in a recent review,[114] "there is no specific conclusion about interaction between coating and alkali-metal atoms" even if good results are obtained with the use of chains of alkanes (such as paraffin[105] that unfortunately works with a maximum operating temperature of $80\,^{\circ}C$), of alkenes,[84] or of organochlorosilanes [such as octadecyltrichlorosilane: OTS (with $E_a = 0.42$ eV[115]) that works only up to $170\,^{\circ}C$[116] but to $350\,^{\circ}C$ with special methods[115]]. Coating the inside of a collimating tube with such materials thus represents a very promising route toward a bright-wall Cs oven but is experimentally non-trivial and was beyond the scope of this study.

### 3. Temperatures effects

Having working-temperature limitations could, however, be hindering the application of these coatings for this purpose because (for instance, from BET theory) it is clear that heating the tube as much as possible seems a good solution to create a bright-wall oven. Indeed, thermal desorption spectroscopy (on diamond)[96] indicates that above $20\,^{\circ}C$, multilayered Cs (or the Cs islands) starts to desorb; then, $100\,^{\circ}C$ is required to desorb Cs from the surface, and $600\,^{\circ}C$ is required to desorb from the defect or grain boundaries. Similarly, at small coverage, alkali metals bound to oxide surfaces (such as $SiO_2$ or $Al_2O_3$) by a strong chemisorption-ionic bond ($E_a = 2.73$ eV), and desorption of the alkali metal occurs between 600 and 1000 K. As the amount of Cs deposited increased, there is a reduction in the Cs desorption temperature to 350 K, reflecting a continuous decrease in the Cs adsorption energy to 0.8 eV with increasing Cs coverage.[117–119] Finally, melting most of the compound formed from the reaction of Cs with air that could have been formed due to an improper pumping is possible at high temperature (CsH at $170\,^{\circ}C$, CsOH at $272\,^{\circ}C$, and $Cs_2O$ and $CsNO_3$ below $610\,^{\circ}C$[90]). Thus, being able to heat the tube material to high-T looks like an important criterion to minimize Cs sticking.

### 4. Test of different materials for the tube

Based on all these observations, we have tested several metals, polymers, glass, or oxides for the tube-materials (typically 2 mm diameter and 80 mm long) such as silver, nickel, Inconel, stainless-steel, copper, OFHC copper, zirconium, Pyrex, TPX, and $Al_2O_3$ with the possibility to heat up to $1000\,^{\circ}C$, thanks to an heated ceramic receptacle surrounding the tube. We did not observe any modification of the output behavior with the tube temperature even when the tube was heated up to $1000\,^{\circ}C$. The only relevant observation was the well known one that we have to heat the tube few tens of degrees above the oven to avoid clogging.

With some materials such as with a copper tube [as shown in Fig. 3(a)], we sometimes indeed succeeded to produce a brightwall oven. However, this behavior holds only for a few days and was impossible to reproduce afterward. In conclusion, testing other tube-materials produced only disappointing results, dominated by dark-wall behavior, except in some very rare and non-reproducible cases. Some of the results are presented in Fig. 4. We found that for all ovens, the theoretical threshold $\lambda \approx d$ is visible experimentally because the beam starts to diverge when entering in the hydrodynamical regime, and this is usually the maximum temperature we use. Figure 4(c) illustrates also clearly that if the divergence is not measured, the estimated brightness can be overestimated easily by a factor 10 (comparing the curve with and without collisions included). This has to be remembered when looking at Fig. 4 and Table I.

Because of the many collisions that would arise from an atom before leaving the tube, a sticking coefficient (ratio of the number of adsorbed atoms that "stick" on the tube surface to the total number of atoms that impinge upon the surface during the same period of time) $S_c < 0.05$ (for our $L/d \sim 50$ case) is required to have a decent bright-wall as simulated using MolFlow (a Monte Carlo simulator for ultrahigh vacuum systems, https://cern.ch/molflow). This means that only a very cleaned surface with a very low sticking coefficient will allow for a bright-wall behavior. Therefore, one of the most probable explanation for the consistent observation of darkwall behavior on all materials is the poor surface quality and/or the surface pollution during manipulations. Our stainless-steel reservoir and our tubes were typically cleaned by tens of minutes in detergent and ultrasonic bath and then acetone and methanol cleaning before baking one hour at $125\,^{\circ}C$ in a furnace. It was then filled with Cs in a glove box filled with argon. However, it is always possible that pollutants from non-pure argon or Cs cells enter the chamber. Clearly, better protocols might be required for better reproducibility, such as cesium distillation to transport pure cesium from a standard Cs crucible to a cleaner oven chamber. Tube surface quality and cleanliness are also probably very important. It has been reported in stainless-steel dosers that chemical reactions occur if surface passivation is not adequate.[120] This passivation might be achieved by flushing the doser overnight with oxygen.[120] We used another approach for the passivation of copper tubes using first 20%–100% nitric acid etch, then 1% aqueous solution of citric acid for several minutes, and then

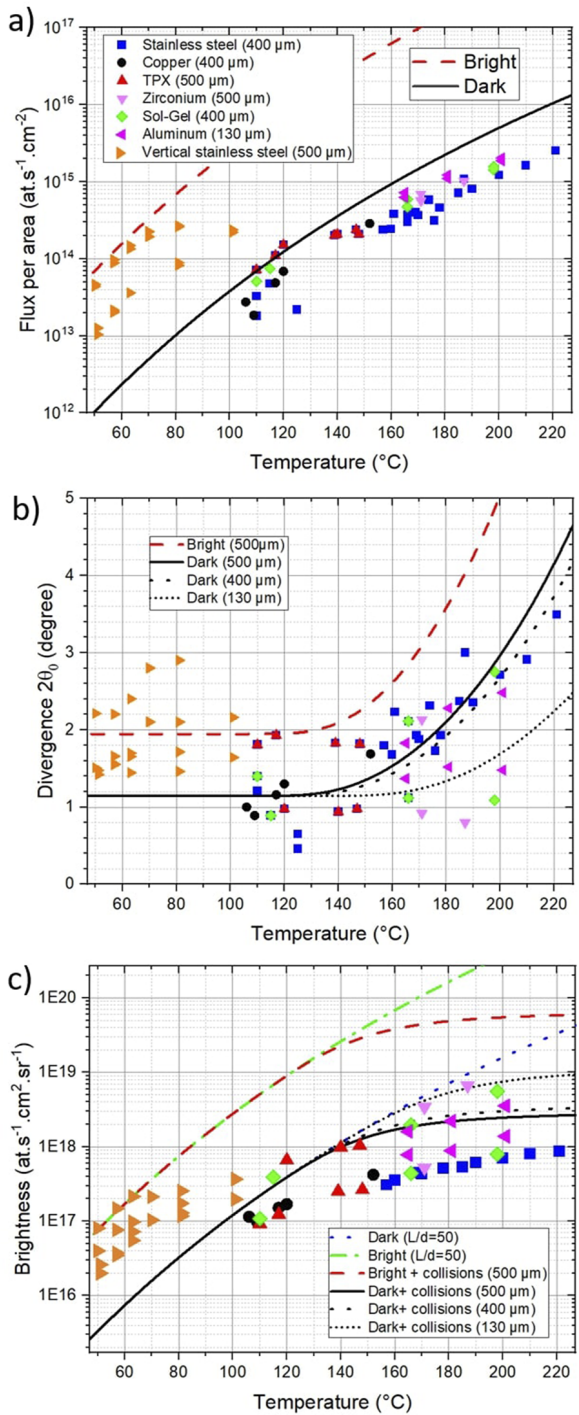

**FIG. 4.** Some of the materials tested for a Cs oven with the design given in Fig. 1(a) and [for the vertical one in Fig. 1(c)]. Different tube diameters $d$ are used (indicated in the figure), but the ratio $L/d = 50$ is constant. (a) Flux per unit area at different temperatures. (b) Divergence of the beam ($2\theta_0$) with the theoretical curve (red dashed curve), taking into account the collisions in the bright-wall case from Eq. (6) and the same theory scaled by the non-collisional factor 1.7 [Eq. (4) over Eq. (5)] to compare with the dark-wall (black dashed curve). (c) Beam brightness for both behaviors.

24 h baking at 100 °C.[121–123] No improvement was observed. This does not mean that such steps are useless but only that other problems occur such as possible exposure to other gases. For instance, depending on the conditioning of the wall (by air-nitrogen or argon exposure), the Cs sticking coefficient was found to be different.[33]

### D. Discussion

The main observation from this study is that it was extremely rare (few days only in few years of operation), and not reproducible, to be in the bright-wall regime, and we were much more often in the dark-wall one. This is a frustrating result mainly because it seems quite limited to cesium, whereas the bright-wall beam seems to be routinely observed for other elements (as a simple example, Rb or Sr has only a factor 4 disagreement with the bright-wall theory[16,124]). For instance, in a dual Li/Cs beam, the Li behavior was the bright-wall one where the Cs flux was found to be a factor 130 times smaller than expected in Ref. 35 [an alternative explanation for this discrepancy (involving worse light-gathering because of Cs coating the windows) is given in the article, but we believe it is instead just another observation that indeed Cs ovens almost always exhibit dark-wall behavior]. It was also clearly observed that following the oven installation and pumping, it takes time (hours for an oven at >120 °C or days if lower temperature) to see any Cs flux out of the tube. We attribute this delay to the necessary time to reach equilibrium between the amount of vapor-phase and adsorbed Cs.[33,78] From our discussion with many colleagues, this behavior seems to be observed in most laboratories and is (naively in the Langmuir's picture) interpreted as the fact that the characteristic time, $\frac{VA\tau}{4\,C}$ where $A$ is the surface area of the cell and $C$ the conductance, to fill stainless-steel systems is very long (and Cs is pumped away from it very slowly).[93] Another interesting observation (although weakened by small statistics) is that the rare cases of bright-wall behavior seem to occur at the early hours of an oven and with quite low heating <100 °C, which then degrades later when we increase the temperature to increase the flux.

All these facts were compatible with the following scenario: for very clean systems and a non-sticking tube, we can observe a bright-wall behavior. However, as time goes by, Cs accumulates on defects, clusters, and reacts to finally lead to the formation of a Cs monolayer on the tube that seems hard to remove even at very high temperature (possibly even forming a stable, complex alloy with the tube substrate). Furthermore, the fact that the new Cs atoms impinging on the tube are not re-emitted even though the tube is very hot (characteristic of a dark-wall oven) indicates that the next layers form a liquid-like phase that migrates by capillarity and hydrodynamics[125] to a colder spot, which is the oven. We have tried to circumvent the layer formation by removing the Cs layer using the light-induced atomic desorption (LIAD) method, sending tens of mW from a 405 nm wavelength laser diode inside the tube, without any success.[126–129]

This phenomenon is compatible with the previously mentioned fact that it takes a long time to observe any Cs flux outside the tube. Indeed, this was always present when using oven geometry of Figs. 1(a) and 1(b) with a vertical stainless-steel oven crucible and a horizontal tube (or apertures) for collimation. In these geometries, the liquid Cs is at the bottom and no direct line of sight exists between the liquid Cs and the oven output, meaning that no Cs atom

**TABLE I.** List of Cs oven properties. The full width divergence angle coefficient $2\theta_0$ is given in degree, the temperature $T$ (of the oven not of the tube or skimmer) is given in °C, the flux $f = \dot{N}$ is given in at. s$^{-1}$, the flux per emitting (nozzle) area $f/A$ is given in at. s$^{-1}$ cm$^{-2}$, the flux per solid angle $f/\Omega$ is given in at. s$^{-1}$ sr$^{-1}$, and the brightness $B = f/A\Omega$ is given in at. s$^{-1}$ cm$^{-2}$ sr$^{-1}$. The information provided by the articles is given in the Roman type, and when no extra information is given, we use the standard formula (and write the result in italic) to extract some values from known data such as $\theta_0 = r/L$ (therefore, we use the dark-wall formula), $\Omega = \pi\theta_0^2$ (for simplicity, we keep this formula even for large $\theta_0$), or $A = \pi r^2$ to calculate $f/A$ or $f/\Omega$. For capillaries, the divergence is chosen from individual one, but the emitting area is the sum over all capillaries (assuming 100% opening ratio, if no extra information is provided). More details are given in the Appendix.

| Year, Reference | Type: Tube material | $2r$(cm) | $L$(cm) | $2\theta_0$ | $T$ | flux $f = \dot{N}$ | $f/A$ | $f/\Omega$ | $B = f/(A\Omega)$ |
|---|---|---|---|---|---|---|---|---|---|
| 1963[61] [a] | Glass tube | 0.1 | 12.7 | 7.3 | 200 | $1.8\times10^{11}$ | $2.3\times10^{13}$ | $1.4\times10^{14}$ | $1.8\times10^{15}$ |
| 1975[62,63] | Stainless steel wick | 1.8 | 9 | *3.4* | 117 | $5.1\times10^{14}$ | $2.0\times10^{14}$ | $1.6\times10^{16}$ | $6.4\times10^{15}$ |
| 1976[64] | Stainless steel | 0.08 | 0.6 | 18 | 127 | $1.0\times10^{14}$ | $4.4\times10^{17}$ | $1.3\times10^{15}$ | $2.6\times10^{17}$ |
| 1981[65] [b] | Stainless steel baffles | 0.35 | 4.5 | 5 | 240 | $4.2\times10^{16}$ | $4.4\times10^{17}$ | $7.0\times10^{18}$ | $7.3\times10^{19}$ |
| 1982[66–68] [c] | Supersonic, Inconel 600 | 1.2 | 9.2 | 7.4 | 457 | $3.9\times10^{20}$ | $3.4\times10^{20}$ | $2.9\times10^{22}$ | $2.6\times10^{22}$ |
| 1985[69–71] [d] | Supersonic, copper | 0.62 | | *17.2* | 447 | $7.6\times10^{20}$ | $1.2\times10^{19}$ | $1.1\times10^{22}$ | $3.6\times10^{22}$ |
| 1985[43] [e] | Porous tube, tungsten | 0.2 | 4.5 | 3 | 132 | $2.2\times10^{13}$ | $7.0\times10^{14}$ | $4.0\times10^{15}$ | $1.3\times10^{17}$ |
| 1986[72] [f] | Glass capillaries + slits | 0.001 | 0.05 | 2.3 | | $1.3\times10^{15}$ | $1.0\times10^{15}$ | $1.0\times10^{18}$ | $8.0\times10^{17}$ |
| 1987[73] [g] | Stainless steel | 0.16 | 10 | *0.9* | 225 | $8.0\times10^{17}$ | $4.0\times10^{18}$ | $1.8\times10^{18}$ | $2.0\times10^{20}$ |
| 1991[74] [h] | Recirculating | 0.15 | 3 | *2.9* | 270 | $3.2\times10^{15}$ | $1.8\times10^{17}$ | $1.6\times10^{18}$ | $9.1\times10^{19}$ |
| 1993[75] [i] | Glass capillaries | 0.001 | 0.05 | 5 | | $8.8\times10^{14}$ | $7.0\times10^{14}$ | $1.5\times10^{17}$ | $1.2\times10^{17}$ |
| 1993[75] [i] | Capillaries + slits | 0.001 | 2.5 | 2.7 | | $1.0\times10^{14}$ | $1.0\times10^{14}$ | $5.6\times10^{16}$ | $5.6\times10^{16}$ |
| 1994[38] [j] | 7 stainless steel tubes | 0.07 | 2 | 15 | 150 | $1.4\times10^{13}$ | $5.3\times10^{14}$ | $2.7\times10^{14}$ | $9.9\times10^{15}$ |
| 1994[38] [j] | Glass capillary array | 0.005 | 0.2 | 5 | 140 | $1.6\times10^{13}$ | $5.1\times10^{14}$ | $2.7\times10^{15}$ | $8.5\times10^{16}$ |
| 2003[76] [k] | 21 steel tubes + glass slits | 0.058 | 1 | 4.1 | 170 | $1.6\times10^{18}$ | $2.1\times10^{18}$ | $3.9\times10^{20}$ | $5.1\times10^{21}$ |
| 2007[77] [l] | Candlestick + aperture | 0.125 | 2.8 | 2.6 | 250 | $3.1\times10^{13}$ | $2.5\times10^{15}$ | $2.0\times10^{16}$ | $1.6\times10^{18}$ |
| 2011[78] [m] | Stainless steel tube | 0.2 | 6 | 60 | 150 | $1.5\times10^{16}$ | $4.7\times10^{17}$ | $1.7\times10^{16}$ | $5.4\times10^{17}$ |
| 2018[35] [n] | 15 stainless steel tubes | 0.051 | 2 | 4.3 | 133 | $2.2\times10^{12}$ | $2.4\times10^{13}$ | $7.4\times10^{15}$ | $7.9\times10^{16}$ |
| 2018[79] [o] | 8 stainless steel waves | 0.022 | 0.9 | *0.7* | 115 | $3.5\times10^{13}$ | $1.4\times10^{15}$ | $3.0\times10^{17}$ | $1.2\times10^{19}$ |
| Figure 3 | Copper | 0.5 | 7 | *4.1* | 85 | $8.0\times10^{14}$ | $4.1\times10^{15}$ | $2.0\times10^{17}$ | $1.0\times10^{18}$ |
| Figure 4 | Vertical stainless steel | 0.05 | 2.5 | 1.6 | 100 | $4.7\times10^{11}$ | $2.3\times10^{14}$ | $7.7\times10^{14}$ | $3.9\times10^{17}$ |
| Figure 4 | Stainless steel | 0.04 | 2.0 | 3.4 | 220 | $3.1\times10^{12}$ | $2.5\times10^{15}$ | $1.1\times10^{15}$ | $9.0\times10^{17}$ |
| Figure 4 | Aluminum | 0.013 | 0.65 | 1.6 | 200 | $2.7\times10^{11}$ | $2.0\times10^{15}$ | $4.3\times10^{14}$ | $3.3\times10^{18}$ |

[a] $L$ 1 in. capillaries (probably in glass) separated by 3 in.; $\theta_0$ calculated from the 3 mm beam size at 7 in. of the 1 mm capillary; and so with a virtual source at $4 \times 7/3$ of an inch.

[b] We take the size from the average size of the extreme holes of 3 and 4 mm. compatible with the divergence found (and used here) in the data at 175 °C.

[c] All references present similar designs and compatible results, so we group them and use Table 1 from Ref. 67 with the condensing skimmer because the classical skimmer creates more diverging flux, however, with a similar brightness.

[d] All references present similar results, so we group them using values from the abstract of Ref. 71. The geometry is complex with nozzle, skimmer, and aperture similar in size, so we do not give any length. We use virtual source a diameter of equivalent area than the nozzle throat.

[e] Higher temperature was used, but above 405 K, $f/\Omega$ starts to saturate. The same oven was studied in Ref. 80, and a theory of a similar oven (with similar values) is given in Ref. 81.

[f] Emitting area $A = 1.25$ cm$^2$, the divergence is measured only along one axis, and no temperature is given. The same system as studied later in Ref. 75.

[g] Diameter and length are taken from Fig. 1. Therefore, the divergence is very approximate.

[h] 1 mm nozzle and 2 mm collimator. Therefore, we choose the middle value. We extract the flux $f = n\bar{v}\pi\theta_0^2 R^2$ from the density $n = 2 \times 10^9$ cm$^{-3}$ at $R = 1.6$ m with a velocity of $\bar{v} = 315$ m/s (assuming Cs velocity due to the tube heated at 350 °C).

[i] Similar system as in Ref. 72. Two measurements exist: In the above line, we give data taken at the location between capillaries and slits with emitting area $A = 1.25$ cm$^2$. In the lower line, data are taken after the slits.

[j] Two measurements: one with seven stainless-steel tubes (first line) and one with a glass capillary array of a total of 2 mm diameter emitting area (lower line). We estimate the brightness from the peak values recorded by the 7 mm hot wire after $R = 1$ m propagation. The measured divergence gives the beam size $\pi\theta_0^2 R^2$ and so, by area ratio, the total current. The highest temperatures lead to more divergence but better brightness; therefore, we choose these values.

[k] The 21 stainless-steel tubes of 1 cm long and 0.023in. diameter (therefore, we put them in the capillary list) should lead to $2\theta_0 = 58$ mrad compatible with the beam size measured after glass slits (that limits the horizontal direction to 13.6 mrad). We thus choose the averaged value for the divergence, but we keep the emitting surface from the tubes of $A = 5 \times 15$ mm$^2$. The flux is estimated by $n\bar{v} \times 13 \times 15$ mm$^2$ from the given density of $n = 3 \times 10^{13}$ cm$^{-3}$ with $\bar{v} = 265$ m/s at 170 °C (110 °C is also given but not really compatible with such a high density).

[l] We used an averaged diameter for the nozzle and aperture. This leads to a divergence that we used because of being compatible with the measured beam size.

[m] Higher flux exists at higher temperature but with a much bigger divergence. The length is chosen to be the nozzle one, but for the diameter, we choose the smaller final aperture. Similar designs are in Refs. 24,25 where the bright-wall theory agrees with experiment.

[n] We assume a perfect circle packing of the 15 tubes (therefore we put this oven in the capillary list) in the equilateral triangle and so an area $A = 56.908 \times (0.81 \text{ mm})^2/2$ for the emitting area.[85]

[o] A source area of $4 \times 2$ mm$^2$ is also mentioned but does not seem to be compatible with the rest, so we keep $A = 4 \times 0.6$ mm$^2$. The capillary collimator can be seen as 64 individual isosceles triangles with 0.5 mm base and 0.15 mm height, leading to quite a different divergence along both directions. However, we choose a divergence given by a diameter to have the same individual area of $A/64$.

emitted from the liquid can directly exit the oven. Thus, the only way a Cs atom can be emitted toward the tube is to be emitted either from the walls in the direction of its axis (meaning that desorption from the stainless-steel walls must be efficient) or from collisions in the gas phase, meaning that the mean free path is small and the oven pressure is high.

We have tested this hypothesis by putting cleaned heated tubes vertically, right above the liquid cesium in the oven, and so in direct view of the liquid as shown in Fig. 1(c). Thus, even at low oven temperature, atoms emitted directly from the surface should be able to traverse the tube and to be re-emitted without ever forming a monolayer on the surface of the tube. In this case, instantaneous (dark or bright-wall) emission should occur at the Cs vapor pressure and so even at quite low temperature. Immediate Cs output was indeed obtained using a stainless-steel vertical tube (the orange points in Fig. 4), confirming the hypothesis. Finally, we note that in such a configuration, as in all bright-wall ones, it was difficult to record higher temperature data simply because the chamber was not pumping efficiently enough, leading to cesium accumulation and a very large background pressure. We have also tested an even simpler configuration of a simple copper aperture ($L = d = 2$ mm) located 7 cm above the liquid cesium. As expected, the emission was instantaneous. However, in this case more interestingly, the divergence of the beam formed was not given by the effusive (small $L/d$ ratio) theory but by the solid angle given by the liquid surface and the distance to the surface aperture. This is simply because no collisions between Cs atoms really occur in the vapor due to the large mean free path at the low temperature (less than 70 °C) used.

## V. CONCLUSION

From a theoretical point of view, a Cs oven with a very small tube that can re-emit atoms (bright-wall) looks ideal for producing high-brightness atomic beams. However, most published experimental studies seem to indicate, on the contrary, that recirculating ovens with slits, following the dark-wall theory, give better results [cf. Fig. 2 (and in Table I)]. The peculiar properties of cesium, namely, its high chemical reactivity and low condensation temperature, are probably the source of this discrepancy. During the course of this study, we tested a lot of different ovens (with various geometries, materials, operating conditions, and cleaning processes), but no clear, reproducible recipe for producing a Cs bright-wall oven could be devised. Bright-wall behaviors have been observed in collimating designs, although very rarely and for about a day, when using a fresh copper tube. Similarly to what is observed when using a non-properly degased SAES Cs dispenser,[130] the properties of a Cs oven depend greatly on the cleanliness of the cesium sample and of the oven itself. However, most, if not all, cesium oven studies (this one included, despite our efforts) do not fulfill these conditions so that surface diffusion (and so formation of atomic structures on defects or aggregations), phase transformations on the walls, or complex mass transfer exists (as observed in Refs. 101, 125, 131).

Better loading systems, for example, using distillation of Cs, might be a way to introduce high-purity cesium properly.[73,132–134] Based on previous experiments and theory, combining this with a better passivated oven crucible and with good (cleaned and without many defects) tube material formed by non-sticking metal or coating or using a porous wall[45,80,135,136] may be promising, as well as arrays of tubes of this type. Vertical designs seem also very promising when the context allows them as low-T bright-wall behavior was observed for a prolonged time (i.e., weeks) on one setup, and Cs emission was found to be immediate.

We hope this study will represent a good starting point for future studies on Cs ovens as there is still a need to better understand their behavior and hopefully obtain reliable conditions for the production of high-brightness atomic Cs beams.

## ACKNOWLEDGMENTS

We acknowledge P. Cheinet and B. Viaris for fruitful discussions and T. Xu for preliminary results. This work was supported by the Fond Unique Interministériel (IAPP-FUI-22) COLDFIB.

## AUTHOR DECLARATIONS

### Conflict of Interest

The authors have no conflicts to disclose.

### DATA AVAILABILITY

The data that support the findings of this study are avalaibe from the corresponding author upon reasonable request.

## APPENDIX: LIST OF Cs OVEN

We give in Table I some characteristics of published Cs ovens. We do not pretend for an exhaustive list of all published Cs ovens, first, because despite our efforts we surely have missed several key articles and, second, because many publications do not focus on the maximum collimating Cs beam capability such as in Refs. 22, 23, 30, 33, 56, 57, 137–151. Most beam measurements are based on the flux, the beam area, or the divergence but rarely, if ever (see Table I), all together. Furthermore, for the area, for the divergence, and so for the beam brightness, we can use the FWHM (this will be our default choice for a diameter when exists), the average one or any other choice depending on the exact beam shape. Our definition of $B = f/A\Omega$ assumes no correlation between the velocity and position; this is roughly the case at the exit of an effusive oven but clearly not the case after a long propagation where at the edge position the lateral velocities are extreme. It is therefore important to trace back the trajectories to get for $A$ the effective source size. Therefore, without any other information, we use for $A$ the emitting area. Similarly, without any extra information about the beam divergence, we will use the dark-wall value for $2\theta_0 = d/L$, which can also lead to an overestimation of the brightness by a factor 3 compared to the bright-wall one, and this is without taking into account collisional effects that enhance the divergence. Therefore, we have to keep in mind that the beam brightness $B$ value we give is probably accurate only within an order of magnitude in most cases.